\documentclass[prl,twocolumn,groupedaddress]{revtex4-1}
\catcode`\@=11
\def\lsim{\mathrel{\mathpalette\@versim<}}
\def\gsim{\mathrel{\mathpalette\@versim>}}
\def\@versim#1#2{\vcenter{\offinterlineskip
\ialign{$\m@th#1\hfil##\hfil$\crcr#2\crcr\sim\crcr } }}
\catcode`\@=12

\catcode`@=12
\usepackage[export]{adjustbox}	
\usepackage{float}
\usepackage{caption}
\usepackage{subcaption}
\usepackage{epsfig,bbm}
\usepackage{amsmath}
\usepackage{slashed}
\usepackage{tikz}
\usepackage{bm}
\usepackage{amssymb}
\usepackage[compat=1.1.0]{tikz-feynman}
\usepackage{mathtools}
\usepackage{graphicx}
\newcommand{\be}{\begin{equation}}
\newcommand{\ee}{\end{equation}}
\newcommand{\bea}{\begin{eqnarray}}
\newcommand{\eea}{\end{eqnarray}}

\begin{document}
\thispagestyle{empty}
\begin{flushright}
\end{flushright}

\begin{center}
\title{ Resolving phenomenological problems with
    strongly-interacting-massive-particle models with dark vector
    resonances
}
\author{Soo-Min Choi}
\email[]{soominchoi90@gmail.com}
\affiliation{Department of Physics, Chung-Ang University, Seoul
06974, Korea}
\author{Pyungwon Ko}
\email[]{pko@kias.re.kr}
\affiliation{School of Physics, Korea Institute for Advanced Study (KIAS),
Seoul 02455, Korea}
\author{Hyun Min Lee}
\email[]{hminlee@cau.ac.kr}
\affiliation{Department of Physics, Chung-Ang University, Seoul
06974, Korea}
\author{Alexander Natale}
\email[]{alexnatale@kias.re.kr}
\affiliation{School of Physics, Korea Institute for Advanced Study (KIAS),
Seoul 02455, Korea}
\begin{abstract}
We consider a light dark matter candidate which is produced by the freeze-out mechanism with $3\rightarrow2$ annihilations, the so called Strongly Interacting Massive Particles (SIMPs). SIMPs are identified as dark pions in dark chiral perturbation theory (ChPT) where both light mass and strong coupling needed for SIMPs can be realized by strong dynamics.  In QCD-like theories with $SU(3)_L \times SU(3)_R /SU(3)_V$ flavor symmetry, including dark vector mesons in the hidden local symmetry scheme, we illustrate that dark vector mesons unitarize the dark ChPT efficiently, thus determine the correct relic density condition within the validity of the dark ChPT.
\end{abstract}
\maketitle
\end{center}

\section{Introduction}

The Weakly Interacting Massive Particles (WIMPs) have been regarded as a consistent paradigm for dark matter (DM), accommodating the large scale structure of the Universe, the gravitational lensing and the Cosmic Microwave Background anisotropies, etc. Nonetheless, deviations from the WIMP paradigm, such as too-big-to-fail ~\cite{Kauffmann:1993gv,Klypin:1999uc,Moore:1999nt,
Bullock:2010uy} and core-cusp problem~\cite{deBlok:2009sp}, have been observed at galaxy scales, rendering the DM profile from the center of the galaxy being cored unlike the expectation of the N-body simulations with WIMPs. These are known as the small-scale problems and would require a large self-scattering cross section for DM in the range of $0.1 {\rm cm^2/g \lesssim \sigma/m_{DM} \lesssim 1 cm^2/g}$~\cite{Tulin:2017ara}. The effects of baryons and
supernova feedback in simulations might resolve such tensions in massive galaxies \cite{Governato:2012fa,Brooks:2012vi}. Otherwise,  the small-scale problems would imply a new long-range force for DM~\cite{Tulin:2017ara} as often appearing in DM models with dark gauge symmetries~\cite{Baek:2013dwa,Ko:2014nha,KO:2016gxk} and/or light dark matter as alternatives to WIMPs. 

Moreover, since there has been no conclusive hint for WIMP signals yet, alternatives to WIMPs have been considered more seriously. Previous direct detection experiments with DM-nucleon elastic scattering don't constrain much the DM masses below $10\,{\rm GeV}$ scale, thus recently there has been more attention to light dark matter candidates. New freeze-out mechanisms for the thermal production of light dark matter with sub-GeV scale mass have been proposed in the literature such as $3\rightarrow 2$ channels~\cite{Carlson:1992fn,Hochberg:2014dra}, forbidden channels~\cite{DAgnolo:2015ujb}, co-decay channels~\cite{Dror:2016rxc} as well as inelastic scattering channels~\cite{DAgnolo:2017dbv}. The last three cases require light hidden particles in the dark sector, whereas the first case with $3\rightarrow 2$ channels rely only on self-interactions of dark matter, thus so called the Strongly Interacting Massive Particles (SIMPs) miracle~\cite{Hochberg:2014dra}.
The freeze-out condition for SIMPs predicts a sub-GeV scale dark matter for solving the small-scale problems, but a consistent model with strong couplings for SIMPs is demanding.

For dark fermions with strong interactions, composite dark pions in the dark Chiral Perturbation Theory (ChPT) are natural candidates for SIMPs~\cite{Hochberg:2014kqa}, as both the DM masses and couplings are derived from strong dynamics. Namely, the dark pion masses are given due to dark QCD condensation while the $3\rightarrow 2$  interactions are naturally obtained by the Wess-Zumino-Witten (WZW) term~\cite{Hochberg:2014kqa}.  However, as the perturbativity limit for 
$m_{\pi}/f_{\pi}$ (i.e. $2 \pi$) is reached by the relic density condition,  the validity of dark ChPT is  prone to considerable NLO and NNLO corrections~\cite{Hansen:2015yaa}.
Indeed, near the validity region of ChPT ($m_\pi / f_\pi , |\vec{p}_\pi| / f_\pi  \sim 2 \pi$), 
there will appear new resonances (as in ordinary hadron physics) 
that have to be included in the chiral Lagrangian explicitly.  Thus, one can never capture correct physics with pions alone near the thresholds of new resonances, thus it is mandatory to include new resonance explicitly in the dark ChPT.

In this letter, we introduce a consistent model for SIMPs in the dark ChPT with QCD-like
chiral symmetry, $SU(3)_L \times SU(3)_R/SU(3)_V$. We regard dark pions as SIMPs and improve the dark ChPT by including vector mesons explicitly as gauge bosons associated with hidden local chiral symmetry. Then, we calculate the thermal relic density 
and the self-scattering cross section for dark pion DM in this framework in  the validity region of the dark ChPT.

\section{Dark chiral Lagrangian  with vector mesons}
The minimal setup of DM scenarios with dark pions are defined in terms of confining non-Abelian gauge theories with the number of dark colors, $N_{h,c}$, and the number of dark flavors, $N_{h,f}$. 
Dark pion DM can be good WIMP candidates with a singlet scalar mediator, as discussed in 
Refs.~\cite{Hur:2007uz,Ko:2008ug,Ko:2009zz,Ko:2010rj,Hur:2011sv,Hatanaka:2016rek}.
A SIMP scenario becomes possible for $N_{h,f} \ge 3$, in which case there exists a nontrivial $\pi_5 (G/H) = Z$ ~\cite{Witten:1983tw,
Witten:1983tx}, thus a nonzero Wess-Zumino-Witten term allows $\pi$-number changing $3\rightarrow 2$ processes.  
In this letter, we consider a dark QCD sector with three flavors of dark quarks in a strongly coupled 
$SU(3)$ gauge symmetry at high energy ($N_{h,c} = N_{h,f} =3$), as proposed in 
Ref.~\cite{Hochberg:2014kqa,Hur:2007uz,Ko:2008ug,Ko:2009zz,Ko:2010rj,Hur:2011sv,Lee:2015gsa}.
(It is straightforward to generalize the numbers of dark colors and dark flavors to different values
~\cite{Hur:2007uz,Ko:2008ug,Ko:2009zz,Ko:2010rj,Hur:2011sv,Hatanaka:2016rek}.)  

The dark sector Lagrangian is given by    
\begin{equation}
\mathcal{L_{SIMP}} = -\frac{1}{4} G^a_{\mu\nu} G^{\mu \nu a} + \sum_{i=1}^{3} \bar{Q}_i \left[  i \slashed{D} - m_i \right] Q_i,
\quad i = 1,2,3.
\end{equation}
In the exact chiral limit $m_i \rightarrow 0$,   this Lagrangian has the exact
global $SU(3)_L \times SU(3)_R$ symmetry, 
which is spontaneously broken into $H_{\rm global} = SU(3)_V$ by nonzero $\bar{Q} Q$ condensate, 
$\langle \bar{Q}_i Q_i\rangle=\Lambda^3_{\rm DQCD}$, with the sum over $i=1,2,3$. 
Then, the broken global symmetry is nonlinearly realized on the coset space $G_{\rm global} /H_{\rm global}$ as 
massless dark pions and  the inclusion of hidden local gauge symmetry as $G_{\rm global} \times H_{\rm local}$ lead 
to dark vector resonances (see Ref.~\cite{Bando:1987br} for review). 

In case of  $m_i \neq 0$, the $SU(3)_V$ will be  explicitly 
broken unless $m_1 = m_2 = m_3$. 
But, $SU(3)_V$ is still an approximately good symmetry as long as $(m_i - m_j )\ll \Lambda_{\rm DQCD}$.  
Also dark pions get masses if $m_i \neq 0$, but they can still be considered as Nambu-Goldstone bosons,
as long as $(m_i + m_j ) \ll \Lambda_{\rm DQCD}$.   In the low energy limit, dynamical degrees of freedom 
in the dark QCD sector will be pseudo Nambu-Goldstone bosons, namely light
dark pions, whose interactions 
are conveniently described by a nonlinear sigma model.  
One can introduce light dark vector mesons (in analogy to the $\rho$ meson in real QCD) as gauge bosons 
associated with $H_{\rm local}$ with the corresponding gauge coupling $g$.  The detailed information on the hidden gauge symmetry and dark vector mesons can be found in the appendix.

The leading chiral Lagrangian for the pion field, $\Sigma(x)={\rm exp}(i2\pi(x)/f_\pi)$, takes the following form,
\begin{eqnarray}
{\cal L}_{\pi} & = & \frac{f_\pi^2}{4} {\rm Tr} \left[ \partial_\mu \Sigma \partial^\mu \Sigma^\dagger \right]. 
\label{pionL}
\end{eqnarray}
Here, we note that $\pi \equiv \pi^a t^a$ fields are normalized such that ${\rm Tr}(\pi^2)=\frac{1}{2}(\pi^a)^2$ with ${\rm Tr} ( t^a t^b ) = \delta^{ab} /2$.
In the presence of extra gauge symmetries, the derivative in the pion Lagrangian (\ref{pionL}) is replaced as $\partial_\mu\Sigma\rightarrow D_\mu\Sigma=\partial_\mu\Sigma-il_\mu \Sigma+i\Sigma r_\mu$ where $SU(3)_L\times SU(3)_R$ is considered as local symmetries.
In this work, however, we set extra gauge symmetry other than dark QCD to zero and keep only the dark pions 
and vector mesons.

Implementing the $SU(3)_V$ as a local symmetry with the  gauge fields, $V_\mu\equiv V^a_\mu t^a$,
we obtain the kinetic term for the vector mesons  as
\be
{\cal L}_{\rm V}= -\frac{1}{2}{\rm Tr}[F_{\mu\nu} F^{\mu\nu}]
\ee
with $F_{\mu\nu}=\partial_\mu V_\nu-\partial_\nu V_\mu-ig[V_\mu,V_\nu]$.

We note that dark pions $\pi(x)$ and vector mesons $V_{\mu}(x)$ are written in the following matrix forms:
\vskip-0.25in
\begin{widetext}
\begin{equation}
\pi (x)  = \frac{1}{\sqrt{2}} ~
\begin{pmatrix}
\frac{1}{\sqrt{2}} \pi^0 + \frac{1}{\sqrt{6}} \eta_8 + \frac{1}{\sqrt{3}} \eta_0 & \pi^+ & K^+ 
\\
\pi^- &  -\frac{1}{\sqrt{2}} \pi^0 + \frac{1}{\sqrt{6}} \eta_8 + \frac{1}{\sqrt{3}} \eta_0 &  K^0 
\\        
K^- & \overline{K^0} & - \frac{2}{\sqrt{6}} \eta_8 + \frac{1}{\sqrt{3}} \eta_0 
\end{pmatrix}
\end{equation}
\begin{equation}
V_\mu(x) = \frac{1}{\sqrt{2}}~
\begin{pmatrix} 
\frac{1}{\sqrt{2}} \rho_\mu^0 + \frac{1}{\sqrt{6}} \omega_{8 \mu}+ \frac{1}{\sqrt{3}} \omega_{ 0 \mu} 
& \rho_\mu^+ & K_\mu^{*+}  
\\
\rho_\mu^- &  -\frac{1}{\sqrt{2}} \rho_\mu^0 + \frac{1}{\sqrt{6}} \omega_{8 \mu} 
+ \frac{1}{\sqrt{3}} \omega_{0 \mu} &  K_\mu^{*0} 
\\        
K_\mu^{*-} & \overline{K_\mu^{*0}} & - \frac{2}{\sqrt{6}} \omega_{8 \mu} 
+ \frac{1}{\sqrt{3}} \omega_{0 \mu} 
\end{pmatrix} 
\end{equation}
\end{widetext}

In terms of the $\Sigma (x)$ field, the pion mass terms are expressed as 
\begin{equation}
{\cal L}_m  =  - \frac{f_\pi^2}{2} {\rm Tr} \left[ \mu (M\Sigma +\Sigma^\dagger M) \right] 
\end{equation}
where $M={\rm diag} (m_1 , m_2 , m_3 )$.
The parameter $\mu$ can be considered as a spurion field, transforming as $\mu \rightarrow 
L^\dagger  \mu R = R^\dagger \mu L$ under $SU(3)_L \times SU(3)_R$ when one constructs the chiral invariant Lagrangian. 
This $\mu$ term breaks chiral symmetry explicitly, thereby generating nonzero masses:
\begin{align}
m_\pi^2 & =  \mu ( m_1 + m_2 )\,,
\\
m_{K^{\pm}} & = \mu ( m_1 + m_3 )\,,
\\
m_{K^0}^2 & =  \mu ( m_2 + m_3 )\,,
\\
m_{\eta_8}^2 & =  \frac{2}{3} \mu (m_1 + m_2 + m_3 )\,,
\\
m_{\eta_0}^2 & =  \frac{1}{3} \mu (m_1 + m_2 + 4 m_3) +  ( 2 \pi  \Lambda)^2. 
\end{align}
Here, $\Lambda^2$ is a correction to the isospin limit, being proportional to $(m_1-m_2)^2$.
For simplicity, we assume $m_1 = m_2 = m_3 \equiv m$ for unbroken remaining $SU(3)_V$ flavor symmetry. 
In this case, all the dark pions have the same mass, $m_{\pi}^2 = 2 \mu m$.  Additionally,
we assume that the vector mesons are also all degenerate with mass $m_V$, as will be discussed shortly.

The resulting Lagrangian is nothing but the usual nonlinear $\sigma$-model Lagrangian. Expanding $\Sigma(x)$ up to trilinear terms in $\pi(x)$, we find that the chiral Lagrangian yields:
\begin{align}
\begin{split}
{\cal L}_\pi &\supset {\rm Tr}[(\partial_{\mu} \pi)(\partial^{\mu} \pi)] \\
&+ \frac{2}{3 f_{\pi}^2} {\rm Tr}[(\partial_{\mu} \pi) \pi (\partial^{\mu} \pi) \pi 
- \pi^2 (\partial_{\mu} \pi) (\partial^{\mu} \pi))]. 
\end{split}
\end{align}
On the other hand, the masses and couplings of vector mesons  are given by
\bea
\Delta {\cal L}_V  &=&  m_V^2 {\rm Tr} V_\mu V^\mu - 2 i g_{V\pi\pi} {\rm Tr} \left( V_\mu [ \partial^\mu \pi , \pi ]  \right) \nonumber \\
&&-\frac{a}{4f^2_\pi} {\rm Tr} \left([\pi,\partial_\mu \pi]^2
\right)
\eea
with
\bea
m_V^2  &=& a g^2 f_\pi^2, \label{EqMV} \\
g_{V\pi\pi}  &=&  \frac{1}{2} a g.  \label{EqG}
\eea
In the ordinary hadron system $a \simeq 2$, but $a$ can be considered as a free parameter 
in the dark ChPT.  In particular we can control $m_V$ and $m_\pi$  independently
by suitably varying the current quark mass $M$.    This is possible, because of  
$m_\pi^2 \sim M \Lambda_{\rm DQCD}$ whereas $m_V^2 \sim \Lambda_{\rm DQCD}^2$.

An important constraint on our model stems from the $2\rightarrow 2$ self-scattering cross section.  The Bullet Cluster
constraints place an upper limit as $\sigma_{scat.}/m_\pi\lesssim 1\,{\rm cm^2/g}$~\cite{Tulin:2017ara}.  
In our model the $2\rightarrow 2$ self-scattering cross section can be calculated for non-relativistic dark pions with relative velocity $v$ by the ChPT Lagrangian to be    
\begin{equation}
\begin{aligned}
\sigma_{scat.} &= 
\frac{77 m_{\pi}^2 }{24 \pi f_{\pi}^4 N_{\pi}^2 } +\frac{m_\pi^2(139f_\pi^2g^2-216m_\pi^2)}{96\pi f_\pi^6 g^2 N_\pi^2 }\,v^2 \\
&+\frac{m_\pi^2}{12288\pi f_\pi^4N_\pi^2}\bigg[ 2176 +\frac{12A_1}{f_\pi^2g^2m_V^2(4m_\pi^2-m_V^2)} \\
&+\frac{9A_2}{f_\pi^4g^4(4m_\pi^2-m_V^2)^2}\bigg] v^4,
\end{aligned}
 \label{EQ2to2a}
\end{equation}
with
\begin{equation}
\begin{aligned}
 A_1 =&\ 6144m_\pi^6-5376m_\pi^4m_V^2+796m_\pi^2m_V^4-7m_V^6, \\ 
A_2 =&\ 10240m_\pi^8-6144m_\pi^6m_V^2+1008m_\pi^4m_V^4\\
&+72m_\pi^2m_V^6-9m_V^8.
\end{aligned}
\end{equation}
The above result becomes the same for $v=0$ as in the dark ChPT  without vector mesons in Ref.~\cite{Hochberg:2014kqa}, being consistent with the low energy theorem \cite{Bando:1987br}. (We note that $N_{\pi}=8$ in our case and the pion decay constant $f_\pi$ in this paper 
is a factor of 2 smaller than that in Ref.~\cite{Hochberg:2014kqa}.)
Vector mesons also lead to velocity-dependent terms in the $2\rightarrow 2$ self-scattering but they are much smaller than the $s$-wave contributions at scales of galaxies and galaxy clusters, even near two-pion resonances, so we safely ignored them.

The chiral Lagrangian with vector mesons presented above has a fictitious symmetry under 
$\pi \rightarrow - \pi$, which is not a true symmetry of dark QCD.  This fictitious symmetry 
would be broken by the 
Wess-Zumino-Witten term~\cite{Wess:1971yu,Witten:1983tw,Witten:1983tx}, 
allowing the processes such as $3\pi \rightarrow KK$ and 
$\pi^0 \rightarrow 2 \gamma$, etc, in ordinary hadron physics.   Generalization of the WZW 
Lagrangian in the presence of vector mesons {$\rho, \omega$, etc.) was obtained in 
Ref.~\cite{Bando:1987br} (see also Ref.~\cite{Ko:1990kd} for compact review on these subjects).

Keep only the dark pions and vector mesons we obtain the full anomalous Lagrangian,
\begin{equation}
\Gamma^{anom} = \int d^4x \left[ \mathcal{L}_{WZW} 
-15C(c_1 {\cal L}_1  + c_2 {\cal L}_2+c_3{\cal L}_3)  \right] 
\end{equation}
where  the Wess-Zumino-Witten term for pions ~\cite{Wess:1971yu,Witten:1983tw,Witten:1983tx} is written as
\begin{equation}
\mathcal{L}_{\rm WZW} = 
\frac{2 N_c}{15 \pi^2 f_{\pi}^5} \epsilon^{\mu \nu \rho \sigma} ~
{\rm Tr}
[
\pi \partial_{\mu} \pi \partial_{\nu} \pi \partial_{\rho} \pi \partial_{\sigma} \pi
],
\label{EqLWZW}
\end{equation}
and  $C \equiv - i \frac{N_c}{240 \pi^2} $.
From the appendix, it can be shown that the gauged WZW terms, ${\cal L}_{1,2,3}$, contain the interactions between vector mesons and pions,  given from the expansion 
in terms of $\pi$ up to $\mathcal{O}(g/f_{\pi}^3)$, as follows,
\bea
{\cal L}_1 &=&  -\frac{ 4g}{f_{\pi}^3} \epsilon^{\mu\nu\rho\sigma} ~{\rm Tr} [ 
V_{\mu} \partial_{\nu} \pi \partial_{\rho} \pi \partial_{\sigma} \pi]=-{\cal L}_2,  \label{WZW1} \\
{\cal L}_3&=&-\frac{2ig}{f_\pi} \epsilon^{\mu\nu\rho\sigma}\, {\rm Tr}[(\partial_\mu V_\nu)(V_\rho \partial_\sigma\pi-\partial_\rho\pi \,V_\sigma)] \nonumber \\ 
&&-\frac{4g^2}{f_\pi}\epsilon^{\mu\nu\rho\sigma}\, {\rm Tr}[V_\mu V_\nu V_\rho \partial_\sigma \pi].   \label{WZW2}
\eea
These new vector meson terms induce additional $3\rightarrow 2$
processes between the dark pions, as illustrated in  Fig.~\ref{FigFeyn}.
In the following numerical analysis, we will first take  $c_1 - c_2 = -1$  and $c_3=1$ in analogy to the hadronic physics. Then we vary $c_1-c_2$ to other values and study phenomenological consequences, since they could be 
different from ordinary hadronic case in principle.

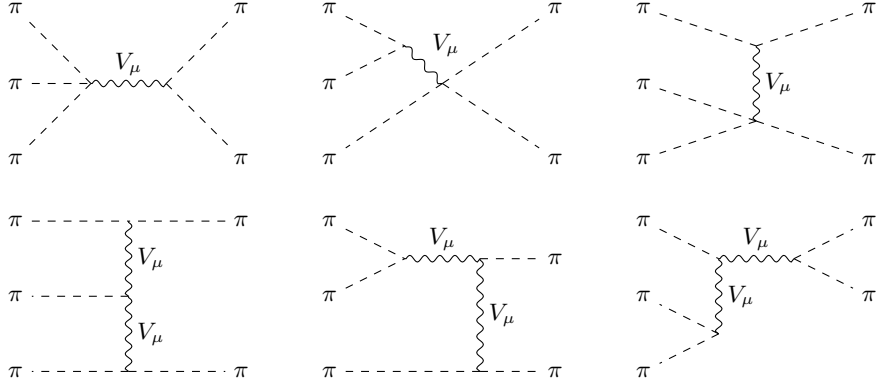
\begin{figure*}[!t]
\begin{center}
\begin{tikzpicture}
\begin{feynman}
    \vertex (i1) {\(\pi\)};
    \vertex at ($(i1) + (0.0cm, -1cm)$) (i2) {\(\pi\)};
    \vertex at ($(i1) + (0.0cm, -2.0cm)$) (i3) {\(\pi\)};
    \vertex at ($(i1) + (3.0cm, 0cm)$) (f1) {\(\pi\)};
    \vertex at ($(i1) + (3.0cm, -2.0cm)$) (f2) {\(\pi\)};
    \vertex at ($(i1) + (1cm, -1cm)$) (m1);
    \vertex at ($(i1) + (2cm, -1cm)$) (m2);
    \diagram* {
      {(i1), (i2), (i3)} -- [scalar] (m1) -- [photon, edge label=\(V_\mu\)] (m2) -- [scalar] {(f1), (f2)}
    };
\end{feynman}
\end{tikzpicture}
\qquad
\begin{tikzpicture}
\begin{feynman}
    \vertex (i1) {\(\pi\)};
    \vertex at ($(i1) + (0.0cm, -1cm)$) (i2) {\(\pi\)};
    \vertex at ($(i1) + (0.0cm, -2.0cm)$) (i3) {\(\pi\)};
    \vertex at ($(i1) + (3.0cm, 0cm)$) (f1) {\(\pi\)};
    \vertex at ($(i1) + (3.0cm, -2.0cm)$) (f2) {\(\pi\)};
    \vertex at ($(i1) + (1cm, -0.5cm)$) (m1);
    \vertex at ($(i1) + (1.5cm, -1cm)$) (m2);
    \diagram* {
      {(i1), (i2)} -- [scalar] (m1) -- [photon, edge label=\(V_\mu\)] (m2) -- [scalar] {(i3), (f1), (f2)}
    };
\end{feynman}
\end{tikzpicture}
\qquad
\begin{tikzpicture}
\begin{feynman}
    \vertex (i1) {\(\pi\)};
    \vertex at ($(i1) + (0.0cm, -1cm)$) (i2) {\(\pi\)};
    \vertex at ($(i1) + (0.0cm, -2.0cm)$) (i3) {\(\pi\)};
    \vertex at ($(i1) + (3.0cm, 0cm)$) (f1) {\(\pi\)};
    \vertex at ($(i1) + (3.0cm, -2.0cm)$) (f2) {\(\pi\)};
    \vertex at ($(i1) + (1.5cm, -0.5cm)$) (m1);
    \vertex at ($(i1) + (1.5cm, -1.5cm)$) (m2);
    \diagram* {
      {(i1), (f1)} -- [scalar] (m1) -- [photon, edge label=\(V_\mu\)] (m2) -- [scalar] {(i2), (i3), (f2)}
    };
\end{feynman}
\end{tikzpicture}
\\
\quad \\
\begin{tikzpicture}
\begin{feynman}
    \vertex (i1) {\(\pi\)};
    \vertex at ($(i1) + (0.0cm, -1cm)$) (i2) {\(\pi\)};
    \vertex at ($(i1) + (0.0cm, -2.0cm)$) (i3) {\(\pi\)};
    \vertex at ($(i1) + (3.0cm, 0cm)$) (f1) {\(\pi\)};
    \vertex at ($(i1) + (3.0cm, -2.0cm)$) (f2) {\(\pi\)};
    \vertex at ($(i1) + (1.5cm, 0cm)$) (m1);
    \vertex at ($(i1) + (1.5cm, -1cm)$) (m2);
    \vertex at ($(i1) + (1.5cm, -2cm)$) (m3);
    \diagram* {
      {(i1), (f1)} -- [scalar] (m1) -- [photon, edge label=\(V_\mu\)] (m2) -- [scalar] {(i2)},
      (m2) -- [photon, edge label=\(V_\mu\)] (m3) -- [scalar] {(i3), (f2)}
    };
\end{feynman}
\end{tikzpicture}
\qquad
\begin{tikzpicture}
\begin{feynman}
    \vertex (i1) {\(\pi\)};
    \vertex at ($(i1) + (0.0cm, -1cm)$) (i2) {\(\pi\)};
    \vertex at ($(i1) + (0.0cm, -2.0cm)$) (i3) {\(\pi\)};
    \vertex at ($(i1) + (3.0cm, -0.5cm)$) (f1) {\(\pi\)};
    \vertex at ($(i1) + (3.0cm, -2.0cm)$) (f2) {\(\pi\)};
    \vertex at ($(i1) + (1cm, -0.5cm)$) (m1);
    \vertex at ($(i1) + (2cm, -0.5cm)$) (m2);
    \vertex at ($(i1) + (2cm, -2cm)$) (m3);
    \diagram* {
      {(i1), (i2)} -- [scalar] (m1) -- [photon, edge label=\(V_\mu\)] (m2) -- [scalar] {(f1)},
      (m2) -- [photon, edge label=\(V_\mu\)] (m3) -- [scalar] {(i3), (f2)}
    };
\end{feynman}
\end{tikzpicture}
\qquad
\begin{tikzpicture}
\begin{feynman}
    \vertex (i1) {\(\pi\)};
    \vertex at ($(i1) + (0.0cm, -1cm)$) (i2) {\(\pi\)};
    \vertex at ($(i1) + (0.0cm, -2.0cm)$) (i3) {\(\pi\)};
    \vertex at ($(i1) + (3.0cm, 0cm)$) (f1) {\(\pi\)};
    \vertex at ($(i1) + (3.0cm, -1cm)$) (f2) {\(\pi\)};
    \vertex at ($(i1) + (1cm, -0.5cm)$) (m1);
    \vertex at ($(i1) + (2cm, -0.5cm)$) (m2);
    \vertex at ($(i1) + (1cm, -1.5cm)$) (m3);
    \diagram* {
      {(i1)} -- [scalar] (m1) -- [photon, edge label=\(V_\mu\)] (m2) -- [scalar] {(f1), (f2)},
      (m1) -- [photon, edge label=\(V_\mu\)] (m3) -- [scalar] {(i2), (i3)}
    };
\end{feynman}
\end{tikzpicture}
\end{center}
\caption{Feynman diagrams contributing to $3\rightarrow 2$ processes for the dark pions 
with the vector meson interactions.}
\label{FigFeyn}
\end{figure*}

\section{Relic density}

In the SIMP scenario,  the $3\rightarrow2$ number-changing processes are assumed to be  
dominant over the DM pair annihilation into SM particles. 
In our model, the total DM density is given by $n_{\rm DM} = \sum_{i=1}^{8} n_i$, and the exact flavor symmetry leads to $N_{\pi}=8$ mass-degenerate dark pions, with $n_1 = n_2 = \cdots = n_8\equiv \frac{1}{8} n_{\rm DM}$. Thus, the resulting Boltzmann equation for $Y_{\rm DM}=n_{\rm DM}/s$ is
\begin{equation}
\frac{d Y_{\rm DM}}{dx} = - \frac{\rho  \langle \sigma v^2 \rangle}{ x^5}(Y_{\rm DM}^3- Y_{\rm DM}^2 Y_{\rm M}^{eq})
\label{ybeq}
\end{equation}	
where $\langle \sigma v^2 \rangle$ is the sum of thermal averaged $3\rightarrow 2$ annihilation cross sections over relevant sub-processes, and  $x \equiv m_{\pi}/T$, and $\rho \equiv 
s^2(m_{\pi}) / H(m_{\pi})$.

\begin{figure*}[!t]
\centering
\begin{minipage}{0.49\textwidth}
\centering
\begin{minipage}{\textwidth}
\centering
\includegraphics[scale=0.53]{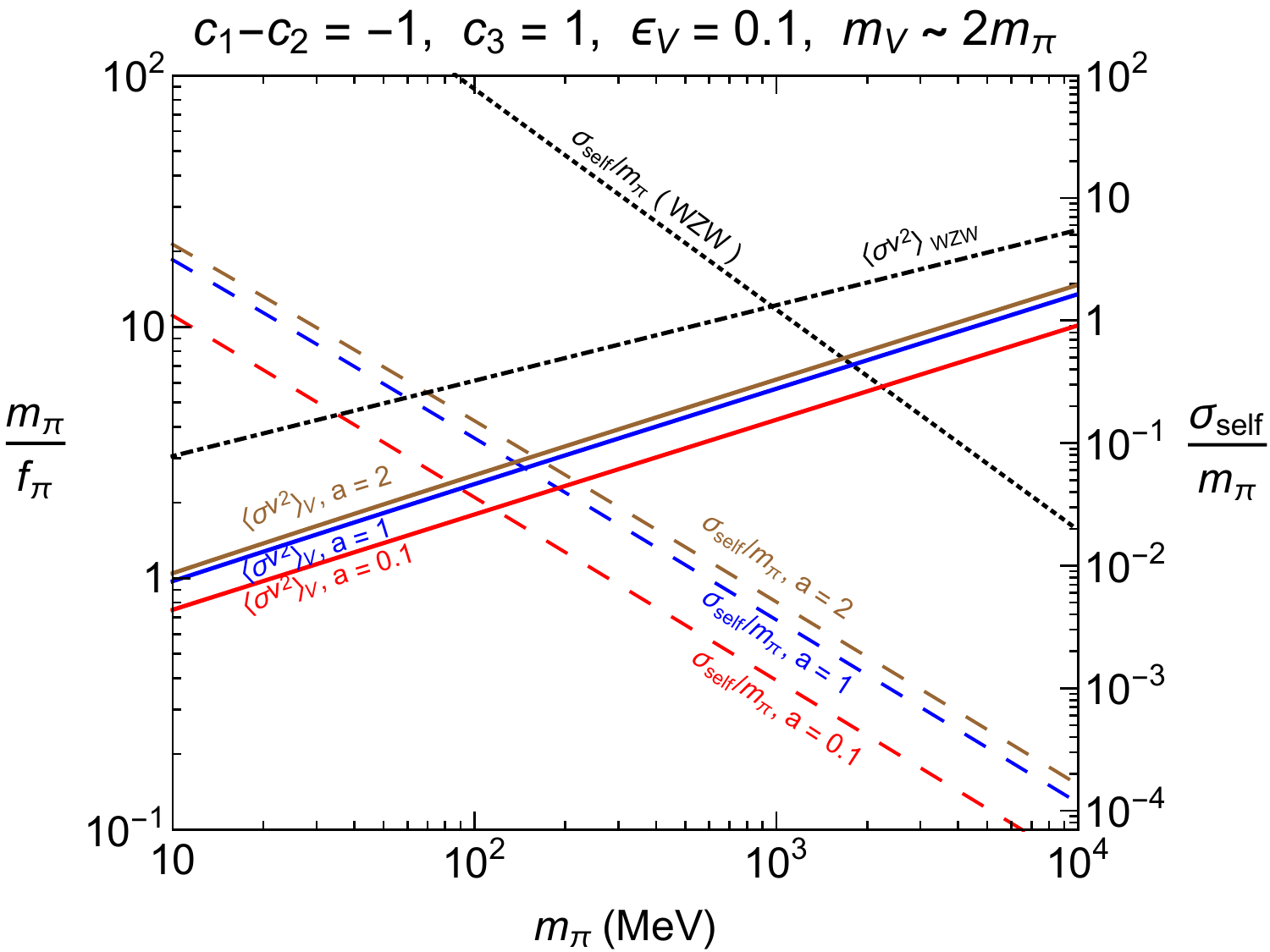}
\end{minipage}
\end{minipage}
\hfill
\begin{minipage}{0.49\textwidth}
\centering
\begin{minipage}{\textwidth}
\centering
\includegraphics[scale=0.53]{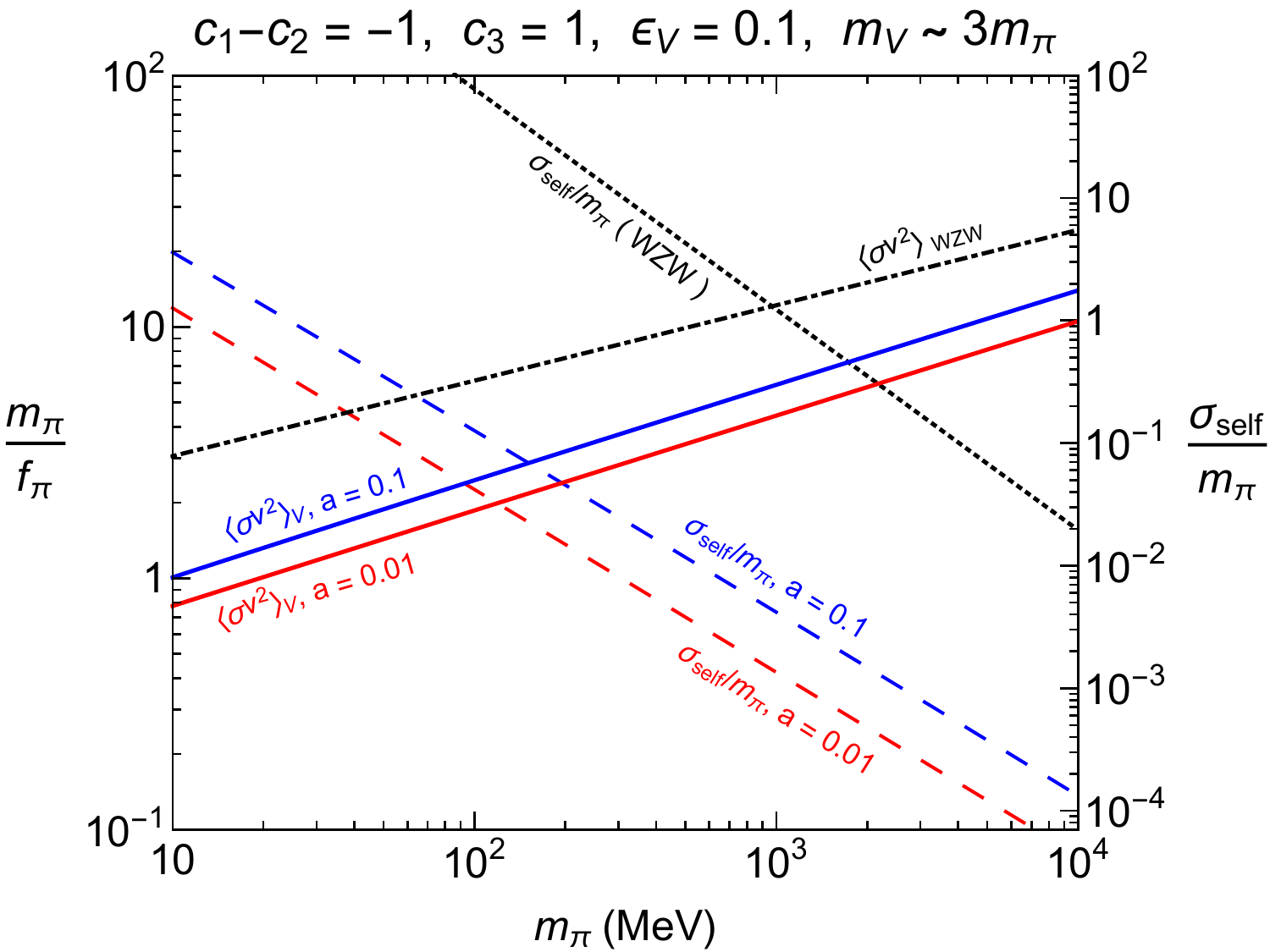}
\end{minipage}
\end{minipage}
\caption{\small Contours of relic density ($\Omega h^2 \approx 0.119$) for $m_\pi$ and $m_\pi/f_\pi$ and self-scattering cross section per DM mass in $\rm cm^2/g$ as a function of $m_\pi$. 
The case without and with vector mesons are shown in black lines and colored lines respectively.
We have imposed the relic density condition for obtaining the contours of self-scattering cross section. 
Vector meson masses are taken near the resonances with $m_{V}= 2(3) m_{\pi}\sqrt{1+\epsilon_V}$ on left(right) plots.  In both plots, $c_1-c_2=-1$ and $\epsilon_V=0.1$ are taken. }
\label{FigRelicSelfScatter}
\end{figure*}

The addition of dark vector mesons in the gauged WZW terms allows for the
$3\rightarrow 2$ processes illustrated in the Fig.~\ref{FigFeyn}, which are not present in the original 
SIMP models \cite{Hochberg:2014dra,Hochberg:2014kqa,Hochberg:2015vrg}, so it can help alleviate the tension between the Bullet Cluster bound and the relic density condition~\cite{Hochberg:2014kqa}.  
Before thermal averaging,  vector resonances modify the $3\rightarrow 2$ annihilation cross section 
in the center of mass frame, as below, 
\bea
( \sigma v^2 )&=&\ \frac{25\sqrt{5}N_c^2m_\pi^5}{\pi^5 f_\pi^{10}N_\pi^3}\Bigg[ \frac{1}{128}
\nonumber\\
&-&\frac{15 a (c_1-c_2) f_\pi^2 g^2 B_1}{1024(4m_\pi^2-m_V^2)(9m_\pi^2-m_V^2)(m_\pi^2+m_V^2)}\nonumber\\
& +&\frac{225 a^2 (c_1-c_2)^2 f_\pi^4 g^4 B_2}{16384(4m_\pi^2-m_V^2)^2(9m_\pi^2-m_V^2)^2(m_\pi^2+m_V^2)^2}\nonumber\\
& -&\frac{225a^3(c_1-c_2)c_3f_\pi^6 g^5 B_3}{8192(4m_\pi^2-m_V^2)^2(9m_\pi^2-m_V^2)^2(m_\pi^2+m_V^2)^3} \label{CS3to2}\\
&+&\frac{15a^2c_3 f_\pi^4g^3 B_4}{512(4m_\pi^2-m_V^2)(9m_\pi^2-m_V^2)(m_\pi^2+m_V^2)^2}\nonumber\\
& +&\frac{225a^4c_3^2f_\pi^8g^6 B_5}{8192(4m_\pi^2-m_V^2)^2(9m_\pi^2-m_V^2)^2(m_\pi^2+m_V^2)^4} \Bigg] b_V\nonumber
\eea
with $b_V =  \frac{1}{4} (v_1^2+v_2^2 +v_3^2)^2-\frac{1}{2}(v_1^4+v_2^4+v_3^4)$ and
\begin{equation}
\begin{aligned}
 B_1 =&\ 37m_\pi^4-21m_\pi^2m_V^2+2m_V^4 \\ 
B_2 =&\ 829m_\pi^8-828m_\pi^6m_V^2+299m_\pi^4m_V^4 \\
&-42m_\pi^2m_V^6+2m_V^8\\
B_3 =&\ 526m_\pi^8 -599 m_\pi^6m_V^2 + 236m_\pi^4m_V^4 \\
&-37m_\pi^2m_V^6+2m_V^8  \\
B_4 = &\ 11m_\pi^4-8m_\pi^2m_V^2+m_V^4\\
B_5 = &\ 170m_\pi^8-218m_\pi^6m_V^2+95m_\pi^4m_V^4\\
&-16m_\pi^2m_V^6+m_V^8.
\end{aligned}
\end{equation}
Here, $v_{1,2,3}$ are the speeds of initial dark pions given in the center of mass frame for the $3\rightarrow 2$ processes, so the corresponding cross section shows a $d$-wave behavior. We note that the $C_V$ term dominates near the resonances at $m_V\approx 2m_\pi$ or $3m_\pi$ for either two pions or three pions.  In these cases, the thermal average should be taken for $2\rightarrow 2$ or $3\rightarrow 2$ Breit-Wigner form \cite{Choi:2016hid,Choi:2017tkj}. 
The width dependence of the $3\rightarrow 2$  cross section is important near resonances, so we included the width of vector mesons explicitly in the vector meson propagators for the later numerical analysis. 

We note that the SIMP scenarios with vector mesons have been studied only in $2\rightarrow 2$ semi-annihilation channels, $\pi\pi\rightarrow \pi V$  \cite{Berlin:2018tvf}. Moreover, the additional channels, $\pi \pi \pi \rightarrow \pi V$ and $\pi \pi \pi\rightarrow VV$, could be relevant for $m_V<2 m_\pi$ and $m_V<3 m_\pi/2$, respectively. But, the latter is closed for $m_V>2m_\pi$; the former  is suppressed by $g_{V \pi \pi}$, which is of order one in the parameter space of our interest, so we didn't include it in our analysis.

\begin{figure*}[!t]
\centering
\begin{minipage}{0.49\textwidth}
\centering
\begin{minipage}{\textwidth}
\centering
\includegraphics[scale=0.53]{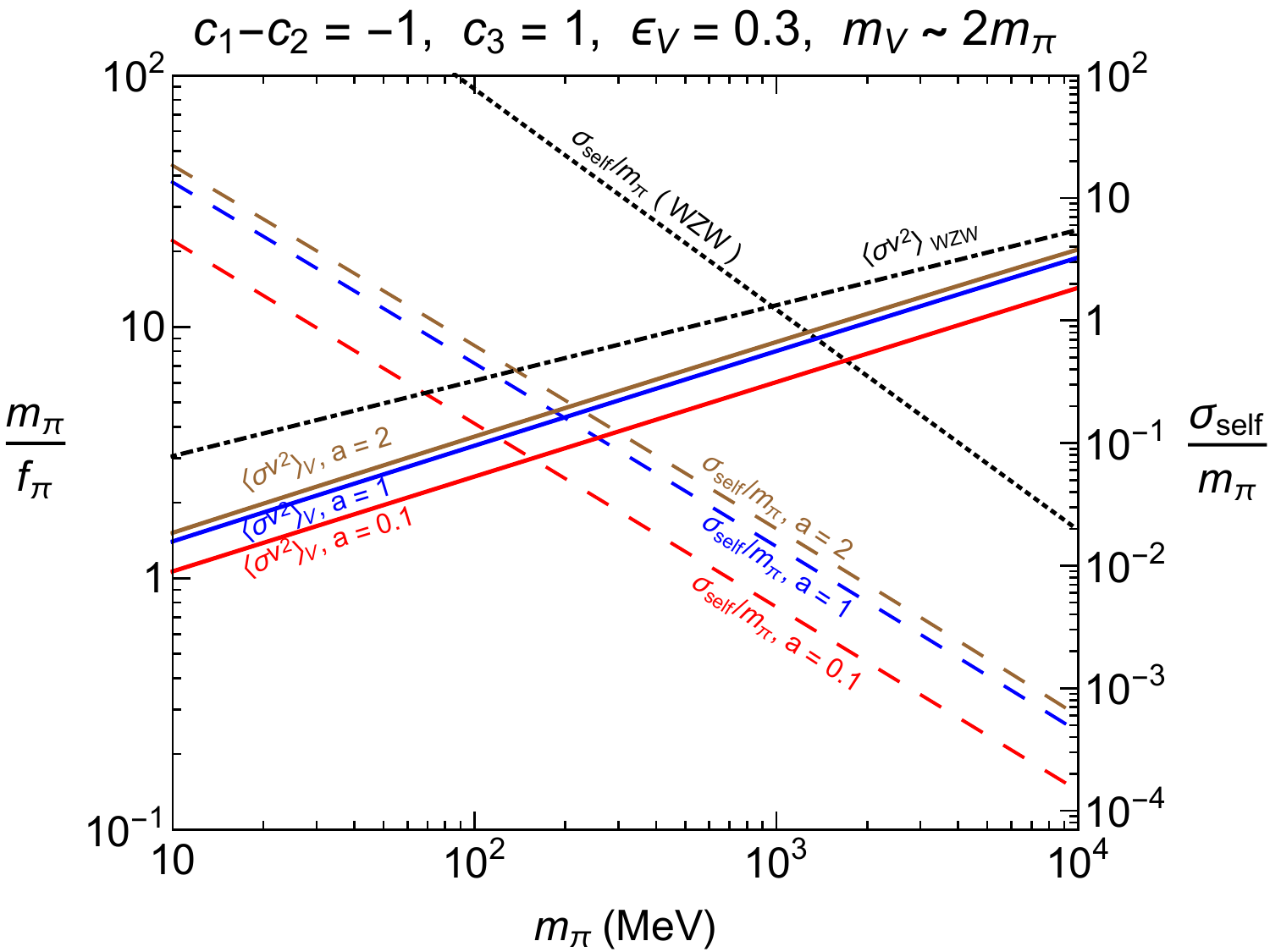}
\end{minipage}
\end{minipage}
\hfill
\begin{minipage}{0.49\textwidth}
\centering
\begin{minipage}{\textwidth}
\centering
\includegraphics[scale=0.53]{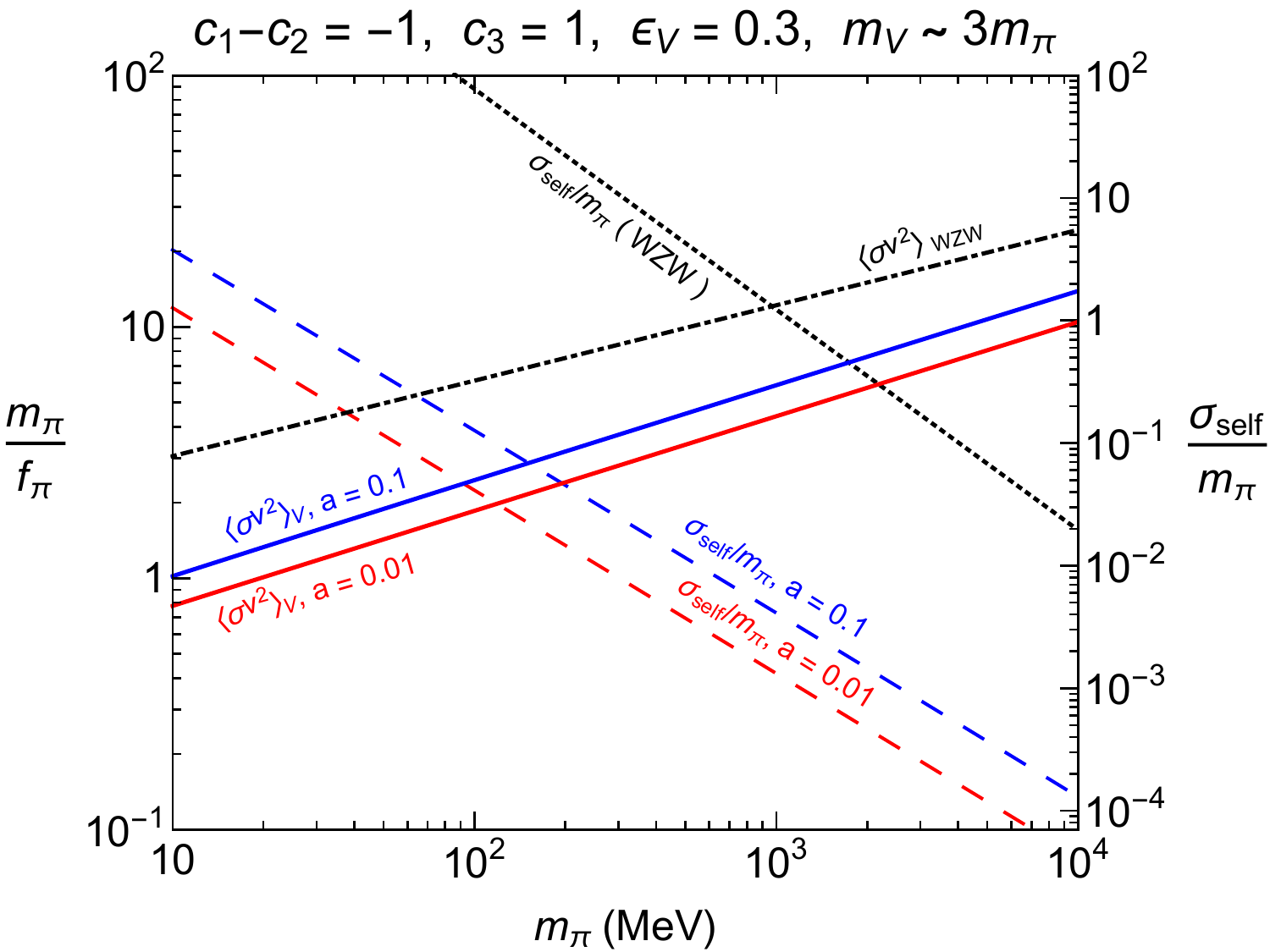}
\end{minipage}
\end{minipage}
\caption{\small Similar contours of relic density for $m_\pi$ and $m_\pi/f_\pi$ and self-scattering cross section per DM mass as in Fig.~\ref{FigRelicSelfScatter}.
Vector meson masses are taken off the resonance with $\epsilon_V=0.3$, and $c_1-c_2=-1$ and $c_3=1$ are chosen.  }
\label{c12-res2}
\end{figure*}

While the $\omega_8$ primarily decays to three pions because $m_\omega <
2 m_K$  in the usual SM QCD,  this is not necessarily true in the case of dark QCD since we can 
vary the pion/kaon mass.  
Since we are assuming all the eight pions/kaons are degenerate in mass, 
two-body decays such as $\omega_8 \rightarrow K K$ could be allowed as well as usual 
three-body decays such as $\omega_8 \rightarrow 3 \pi$. 
Then we find that the widths of vector mesons with degenerate masses are identical as follows,
\begin{equation}
\Gamma_V = \frac{a^2 g^2 m_V}{256 \pi}\left(1-4\frac{m_{\pi}^2}{m_V^2}\right)^{3/2}.
\label{gammaV}
\end{equation}
If we chose a QCD-like set of parameters ($a \approx 2$, $c_1-c_2=-1$ and $c_3=1$), the widths of vector mesons would be sizable for values of $m_{\pi}/f_{\pi}$ that yield the correct relic density.  However, if $a \ll 1$, then the mass relation, $m_V^2 = a g^2 f_{\pi}^2 \approx
9 m_{\pi}^2$ or $4 m_{\pi}^2$, is maintained with $\Gamma_V/m_V \ll 1$.

For $3\rightarrow 2$ processes, we take the vector meson masses near the resonances and make the thermal average under the narrow width approximation with $\Gamma_V/m_V \ll 1$ in Eq.~(\ref{CS3to2}).
Then, the thermal averaged $3\rightarrow 2$ annihilation cross section becomes \cite{Choi:2017tkj}
\begin{equation}
\langle \sigma v^2 \rangle_R \approx \left\{ \begin{array}{c} \frac{81 \pi}{128}\,\kappa \epsilon^4_V x^3 e^{-\frac{3}{2} \epsilon_V x},\,\,\, \quad\quad m_V\approx 3m_\pi, \\ \frac{8}{3}\sqrt{\pi} \, \kappa \epsilon^{3/2}_V x^{1/2}\, e^{-\epsilon_V x},\,\,\, \,\,\, m_V\approx 2m_\pi,  \end{array} \right.
\end{equation}
where the effective $3\rightarrow 2$ cross section before thermal average is taken to be $(\sigma v^2)= \frac{\kappa b_V \gamma_V}{(\epsilon_V-u^2)^2+\gamma^2_V}$, with $\kappa$ being the velocity-independent coefficient, 
$(\epsilon_V,\gamma_V) = (\frac{m_V^2 - 4 m_{\pi}^2}{4 m_{\pi}^2},\frac{m_V \Gamma_V}{4 m^2_\pi})$ and $u^2=\frac{1}{2}(v^2_1+v^2_2)-\frac{1}{4} v^2_3$  for two-pion resonances or $(\epsilon_V,\gamma_V)=(\frac{m_V^2 - 9 m_{\pi}^2}{9m_{\pi}^2},\frac{m_V\Gamma_V}{9m^2_\pi})$ and $u^2=\frac{1}{3}(v^2_1+v^2_2+v^2_3)$ for three-pion resonances. 
Then, we can solve the Boltzmann equation by fixing the vector meson masses or $\epsilon_V$ and find the condition for the correct relic density.

\begin{figure*}[!t]
\centering
\begin{minipage}{0.49\textwidth}
\centering
\begin{minipage}{\textwidth}
\centering
\includegraphics[scale=0.53]{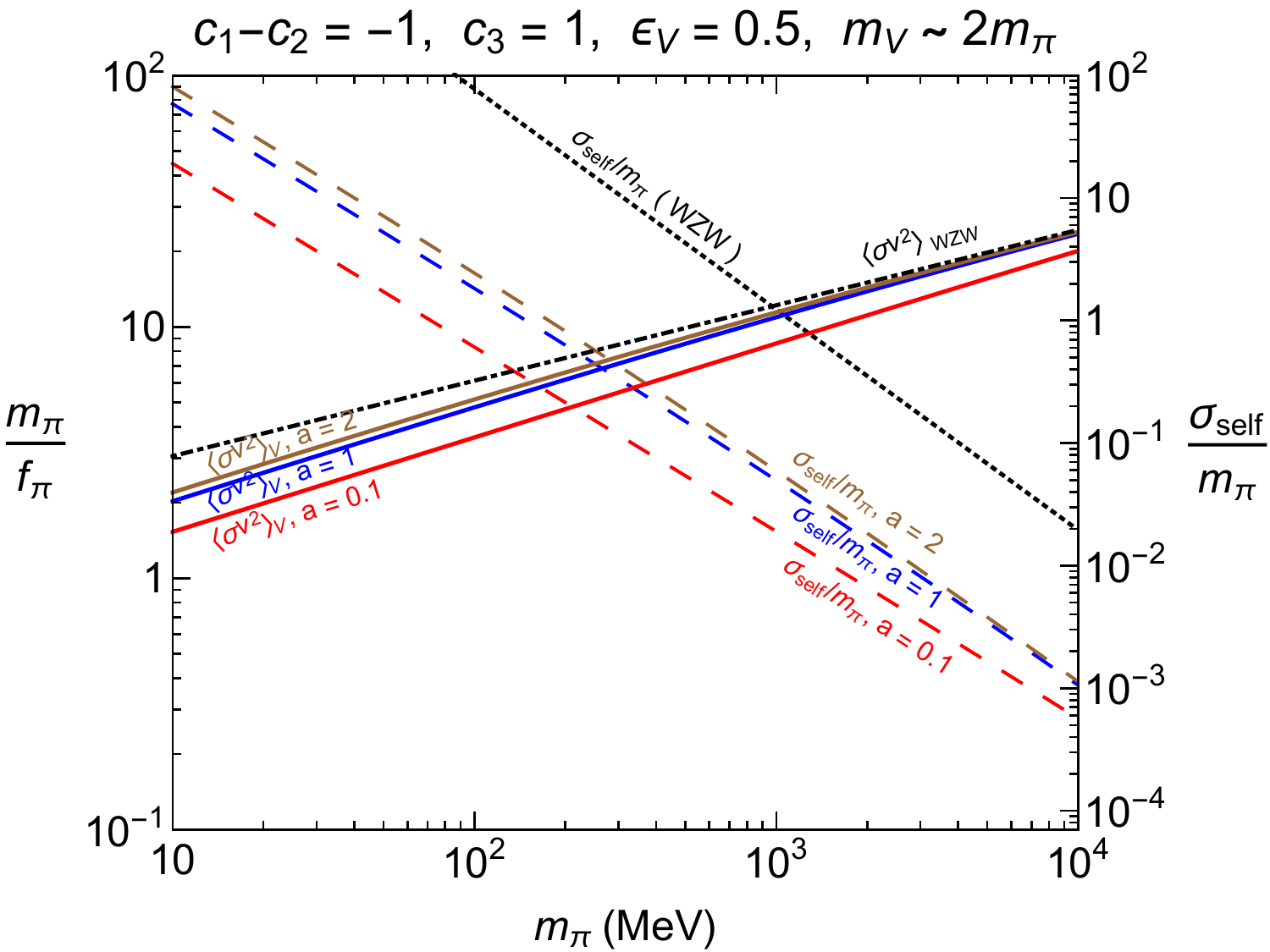}
\end{minipage}
\end{minipage}
\hfill
\begin{minipage}{0.49\textwidth}
\centering
\begin{minipage}{\textwidth}
\centering
\includegraphics[scale=0.53]{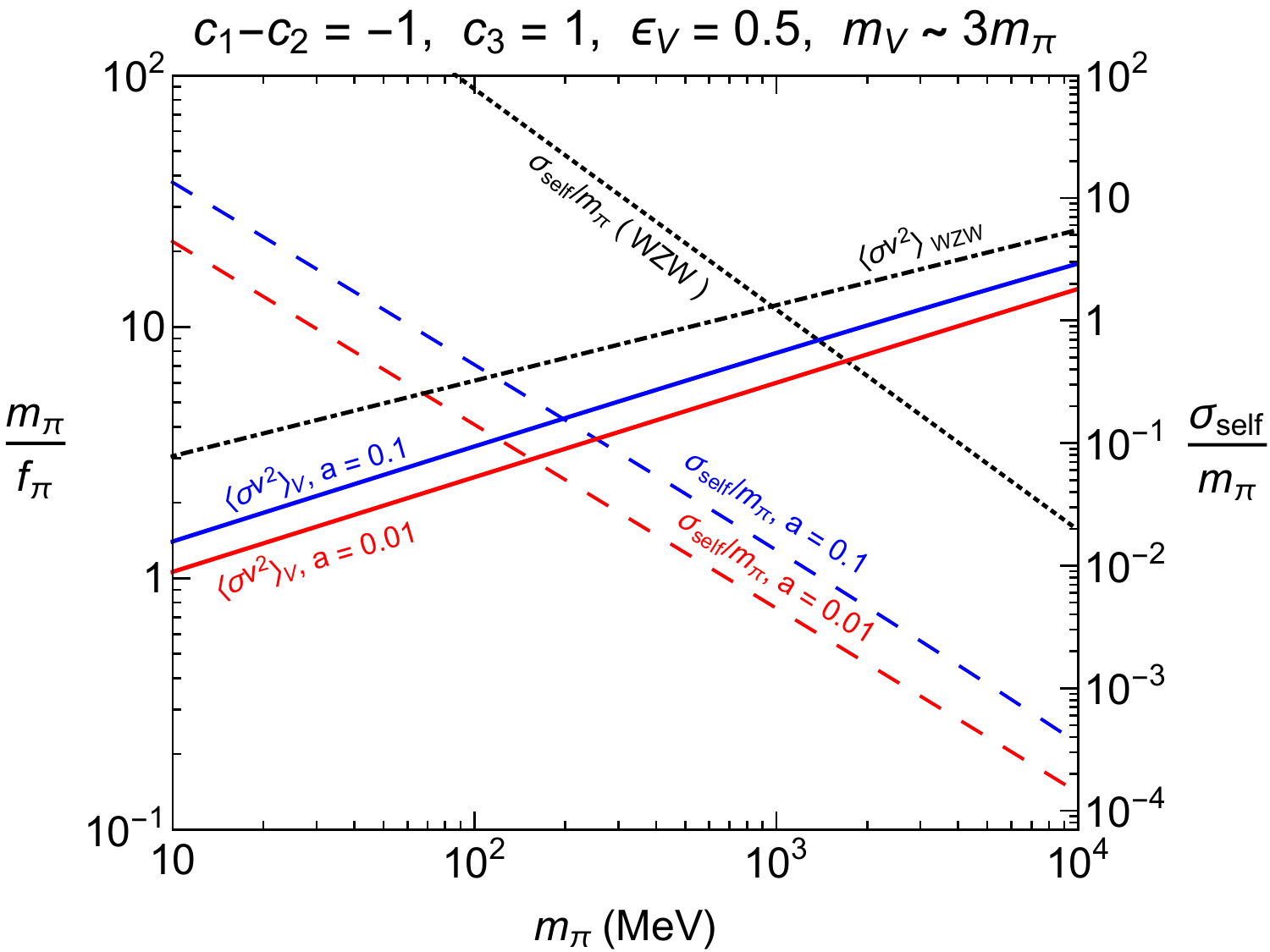}
\end{minipage}
\end{minipage}
\caption{\small Similar contours of relic density for $m_\pi$ and $m_\pi/f_\pi$ and self-scattering cross section per DM mass as in Fig.~\ref{FigRelicSelfScatter}.
Vector meson masses are taken off the resonance with $\epsilon_V=0.5$, and $c_1-c_2=-1$ and $c_3=1$ are chosen. }
\label{c12-res3}
\end{figure*}

In Fig.~\ref{FigRelicSelfScatter}, we illustrate contours of constant relic density 
($\Omega h^2 \approx 0.119$)  for $m_\pi$ vs $m_\pi/f_\pi$ and the dark pion self-scattering cross section as a function of $m_\pi$  for the value of $f_{\pi}$ that yields the correct relic density. 
Parametrizing vector meson masses by $m_{V}= 2(3) m_{\pi}\sqrt{1+\epsilon_V}$ on left(right) plots, we have chosen $c_1-c_2=-1$, $c_3=1$ and $\epsilon_V=0.1$ for both plots in Fig.~\ref{FigRelicSelfScatter}.  
Taking the WZW terms without vector mesons, we show the relic density condition in black dot-dashed lines and the self-scattering cross section without 
vector mesons in black dotted lines in both plots, respectively. 
For different choices of $a$, the relic density condition is satisfied in colored solid lines and 
the corresponding self-scattering cross sections are shown in colored dashed lines.

As can be seen in Fig.~\ref{FigRelicSelfScatter}, the value of $m_{\pi}/f_{\pi}$ needed for the correct relic density is reduced due to vector meson resonances  with  
$a = O(1)$ ($a \ll 1$) for $m_V \sim 2 m_\pi$ ($m_V \sim 3 m_\pi)$, as compared with the case 
with the WZW terms without vector mesons. 
The self-scattering cross section in our scenario with vector mesons is greatly reduced due to a smaller value of $m_\pi/f_\pi$ than in the case without vector mesons.  We have checked that varying the anomalous parameters  $c_{1,2,3}$, acceptable values for the relic density and the self-scattering cross section can be obtained within the validity region of chiral perturbation theory with light vector mesons.

We remark on the vector meson coupling, $g_{V\pi\pi}=\frac{3}{2}(1)\sqrt{a}(m_\pi/f_\pi)\sqrt{1+\epsilon_V}$, near the three(two)-pion resonance, from Eqs.~(\ref{EqMV}) and (\ref{EqG}).  First, for $m_V \sim 3 m_{\pi}$, $c_1-c_2=-1$ and $c_3=1$ (on right in Fig.~\ref{FigRelicSelfScatter}),  the correct relic density requires $m_\pi/f_\pi\lesssim 6(4.5)$ for $a=0.1(0.01)$ and  $m_\pi\lesssim 1\,{\rm GeV}$, but  we need $g_{V\pi\pi}\lesssim 3.0(0.7)$ in this case. For $m_V \sim 2 m_{\pi}$, $c_1-c_2=-1$  and $c_3=1$ (on left in Fig.~\ref{FigRelicSelfScatter}),  the correct relic density requires $m_\pi/f_\pi\lesssim 5.5(4)$ for $a=1(0.1)$ and $m_\pi\lesssim 1\,{\rm GeV}$, resulting in $g_{V\pi\pi}\lesssim 5.8(1.3)$, which is comparable to the case with $m_V \sim 3 m_{\pi}$. Then, the unitarity violation is delayed to much higher energy scales due to vector mesons in our scenario, although not far from the scale of vector meson masses, for instance, through $V\pi\rightarrow \pi\pi$.

Off the resonance poles, there is still a meaningful improvement of perturbativity with vector mesons.
In Figs. 3 and 4, we take the vector meson masses off the resonance poles to $\epsilon_V= 0.3$ and $0.5$ with respect to $m_V = 2 m_\pi$ and $m_V = 3 m_\pi$ on left and right panels, respectively.    With $m_V=3m_\pi \sqrt{1+\epsilon_V}$ and $\epsilon_V=0.5$, the correct relic density requires $m_\pi/f_\pi\lesssim 8(6)$ for $a=0.1(0.01)$ and  $m_\pi\lesssim 1\,{\rm GeV}$, thus  $g_{V\pi\pi}\lesssim 4.6(1.6)$;  with $m_V=2m_\pi \sqrt{1+\epsilon_V}$ and $\epsilon_V=0.3$, the correct relic density requires $m_\pi/f_\pi\lesssim 8(6)$ for $a=1(0.1)$ and  $m_\pi\lesssim 1\,{\rm GeV}$, thus  $g_{V\pi\pi}\lesssim 9(2)$. Therefore, we may tolerate vector meson masses to be further off the resonance conditions, $m_V=2 m_\pi$ or $m_V=3m_\pi$, being consistent with perturbativity and extending a viable parameter space.

Before closing, two remarks are in order.   First of all, if the assumption of degenerate masses is relaxed,  the thermal relic density could be achieved in some interesting parameter space, which we hope to return in a future publication. Secondly, in the SIMP scenario, 
the dark sector is required to remain in kinetic equilibrium with the SM~\cite{Hochberg:2014dra}. This is accomplished via portal interactions for dark scalars such as sigma field (or dark Higgs)
~\cite{Kamada:2016ois,Choi:2017zww} or dark photon
~\cite{Lee:2015gsa,Choi:2015bya,Choi:2016tkj}, 
the details of which would deserve a further study for the detection of SIMP dark matter.

\section{Conclusions}
We have considered a SIMP scenario where dark pions in the dark QCD are light dark matter candidates. Including dark vector mesons in the hidden gauge symmetry scheme, we showed 
that the $3\rightarrow 2$ annihilation cross section can be enhanced near resonance poles  to realize the SIMP freeze-out mechanism, while reducing the self-scattering cross section. As a result, we proposed a consistent scenario for natural light dark matter  with $3\rightarrow 2$ processes where there is no perturbativity problem for the parameter values rendering the correct relic density.

\textit{Acknowledgments}--- 
We would like to thank Eric Kuflik for helpful discussion.
The work of SMC and HML are supported in part by Basic Science Research Program through
the National Research Foundation of Korea (NRF) grant NRF-2016R1A2B4008759. The work of SMC is supported in part by TJ Park Science Fellowship of POSCO TJ Park Foundation.  The work of PK and AN are supported in part by Korea Research Foundation (NRF) Research Grant NRF-2015R1A2A1A05001869. 

\section{Appendix}
Here we provide the details for the chiral Lagrangian with vector mesons for QCD-like chiral symmetry, $SU(3)_L\times SU(3)_R/SU(3)_V$, in the hidden local gauge symmetry scheme. We also list the anomalous WZW Lagrangian that is responsible for four-point interactions between dark pions and vector mesons. 

It is convenient to introduce the fields which transform  under global 
$SU(3)_L \times SU(3)_R$ and local $SU(3)_V$  as follows:
\begin{align}
\xi_L (x) & \rightarrow  U(x) \xi_L (x) L^\dagger
\\
\xi_R (x) & \rightarrow  U(x) \xi_R (x) R^\dagger 
\\
g V_\mu (x) & \rightarrow  U(x) \left[ \partial_\mu - i g V_\mu (x) \right] U^\dagger (x) 
\\
D_\mu \xi_L & =  ( \partial_\mu - i g V_\mu ) \xi_L (x) + i \xi_L (x) l_\mu
\\
D_\mu \xi_R & =  ( \partial_\mu - i g V_\mu ) \xi_R (x) + i \xi_R (x) r_\mu
\end{align}
Here $L \in SU(3)_L$, $R \in SU(3)_R$ and $U(x) \in SU(3)_V$, and we have implemented 
the global $SU(3)_L \times SU(3)_R$ as local symmetries, by introducing $l_\mu$ and $r_\mu$ 
as gauge fields of the local $SU(3)_L \times SU(3)_R$ gauge symmetries and identifying them 
as the gauge bosons of any additional dark gauge symmetries.

Then the chiral Lagrangian for dark pions and vector mesons is given by 
\begin{align}	
{\cal L} & =  {\cal L}_A + {\cal L}_B 
+{\cal L}_m + {\cal L}_{\rm kin} + \Gamma^{\rm anom}
\\
{\cal L}_A & =  - \frac{f_\pi^2}{4} {\rm Tr} \left[ ( D_\mu \xi_L )\xi_L^\dagger - 
(D_\mu \xi_R )\xi_R^\dagger \right]^2 
\label{LA}
\\
{\cal L}_B & =  - a \frac{f_\pi^2}{4} {\rm Tr} \left[ ( D_\mu \xi_L )\xi_L^\dagger + 
(D_\mu \xi_R )\xi_R^\dagger \right]^2 
\label{LB}
\\
{\cal L}_m & =  - \frac{f_\pi^2}{2} {\rm Tr} \left[ \mu (\xi_L M \xi_R^\dagger+ H.c.) \right] 
\label{Lm}
\\
{\cal L}_{\rm kin} & =  - \frac{1}{2} {\rm Tr} \left[ F_{\mu\nu} F^{\mu\nu} \right] 
\label{Lkin}
\\
F_{\mu\nu} & =   \partial_\mu V_\nu - \partial_\nu V_\mu - ig [ V_\mu , V_\nu ]
\end{align}

One can also define a new exponential field $\Sigma (x)$ as 
$\Sigma (x) \equiv \xi_L^\dagger (x) \xi_R (x) = {\rm exp} [ i 2 \pi(x)/f_\pi ]$ 
with $\xi_L^\dagger (x) = \xi_R (x) = {\rm exp} [ i \pi(x)/f_\pi ]$ which transforms as 
$\Sigma (x) \rightarrow L \Sigma (x) R^\dagger$ under the  original global chiral transformation.

We define the following objects,
\begin{align}
\hat{\alpha}_L & =  D\xi_L \cdot \xi_L^\dagger = \alpha_L - i g V + i \hat{l},
\\
\hat{\alpha}_R & =  D\xi_R \cdot \xi_R^\dagger = \alpha_R - i g V + i \hat{r}, 
\\
F_V & =  dV- i g V^2, 
\\
{\hat F}_L & = \xi_L (dl-il^2) \xi_L^\dagger, \\
{\hat F}_R & = \xi_R (dr-ir^2) \xi_R^\dagger
\end{align}
with $\alpha_{L/R} =  d\xi_{L/R} \cdot \xi_{L/R}^\dagger$, ${\hat l}=\xi_L\cdot l\cdot \xi^\dagger_L$ and ${\hat r}=\xi_R\cdot r\cdot \xi^\dagger_R$.
Then, the anomalous WZW terms in the presence of light vector mesons are given by
\begin{align}
\Gamma^{anom} & =  \Gamma_{\rm WZW} -15 C \sum_{i=1}^4 c_i \int d^4x {\cal L}_i
\\
{\cal L}_1 & =  {\rm Tr} \left[  \hat{\alpha}_L^3 \hat{\alpha}_R - 
\hat{\alpha}_R^3 \hat{\alpha}_L  \right] 
\\
{\cal L}_2 & =  {\rm Tr} \left[  \hat{\alpha}_L \hat{\alpha}_R \hat{\alpha}_L \hat{\alpha}_R  \right] 
\\
{\cal L}_3 & =  i {\rm Tr} \left[ F_V ( \hat{\alpha}_L \hat{\alpha}_R -  \hat{\alpha}_R \hat{\alpha}_L
)\right] 
\\
{\cal L}_4 & =  i {\rm Tr} \left[ \hat{F}_L \hat{\alpha}_L \hat{\alpha}_R - \hat{F}_R \hat{\alpha}_R 
\hat{\alpha}_L \right] 
\end{align}
where
\begin{equation}
C = - i \frac{N_c}{240 \pi^2} .
\label{Canom}
\end{equation}
Here, $\Gamma_{WZW}$ is the familiar Wess-Zumino-Witten term for pions
~\cite{Wess:1971yu,Witten:1983tw,Witten:1983tx}, written as 
\begin{equation}
\Gamma_{WZW} = C \int d^5x {\rm Tr} ( \alpha^5 ).
\end{equation}
In this work, there is no extra gauge symmetry other than dark QCD.  Therefore we ignore 
the external gauge fields by setting $l_\mu = r_\mu = 0$ and keep only the dark pions 
and vector mesons $V_\mu$, thus $\mathcal{L}_{4}$ is zero.

\bibliography{Bib}

\end{document}